\documentclass[10pt]{article}

\usepackage{graphicx}
\usepackage{subcaption}

\usepackage[a4paper, left=35mm,right=35mm,top=34mm,bottom=34mm]{geometry}
\usepackage[utf8]{inputenc}
\usepackage[T1]{fontenc}
\usepackage[english]{babel}

\usepackage{enumerate}
\usepackage{graphicx}
\usepackage{hyperref}
\hypersetup{
    colorlinks=true,
    linkcolor=blue,
    filecolor=magenta,      
    urlcolor=cyan,
}

\usepackage{mathtools,amsthm,amssymb,amsfonts}
\usepackage{algorithm}
\usepackage{algpseudocode}
\makeatletter
\def\thm@space@setup{
  \thm@preskip=10pt \thm@postskip=10pt
}
\makeatother
\theoremstyle{plain}

\theoremstyle{plain}

\theoremstyle{definition}

\theoremstyle{definition}

\theoremstyle{remark}

\theoremstyle{remark}
\newtheorem{assumption}{Assumption}

\usepackage{caption} 
\usepackage{booktabs}
\usepackage{listings}
\usepackage{color}
\definecolor{dkgreen}{rgb}{0,0.6,0}
\definecolor{gray}{rgb}{0.5,0.5,0.5}
\definecolor{mauve}{rgb}{0.58,0,0.82}
\usepackage{enumitem}

\newcommand{\abs}[1]{\left\lvert#1\right\lvert}
\newcommand{\norm}[1]{\left\lVert#1\right\rVert}

\newcommand{\email}[1]{\protect\href{mailto:#1}{#1}}

\usepackage{xcolor}[2.11]
\colorlet{inlinkcolor}{green!50!black}
\colorlet{exlinkcolor}{red!50!black}
\if@hidelinks
\hypersetup{ hidelinks = true }
\else
\hypersetup{
  colorlinks = true,
  allcolors = inlinkcolor,
  urlcolor = exlinkcolor,
}
\fi

\newenvironment{@abssec}[1]{
        \vspace{.05in}\parindent .0in
        {\upshape\bfseries #1. }\ignorespaces
    }
    {\par\vspace{.1in}}
\renewenvironment{abstract}{\begin{@abssec}{\abstractname}}{\end{@abssec}}
\newenvironment{keywords}{\begin{@abssec}{Keywords}}{\end{@abssec}}

\usepackage{fancyhdr}

\author{
  {\normalsize Fr\'ed\'eric Magoul\`es}\thanks{CentraleSup\'elec, Universit\'e Paris-Saclay, 3 rue Joliot Curie, 91190 Gif-sur-Yvette, France
    (\email{frederic.magoules@hotmail.com}, \email{zouqinmeng@gmail.com}).}
  \and
  {\normalsize Qinmeng Zou\footnotemark[1]}
}
\title{Asynchronous Time-Parallel Method based on Laplace Transform}
\date{}

\begin{document}
\maketitle
\thispagestyle{fancy}

\begin{abstract}
Laplace transform method has proved to be very efficient and easy to parallelize for the solution of time-dependent problems.
However, the synchronization delay among processors implies an upper bound on the expectable acceleration factor, which leads to a lot of wasted time.
In this paper, we propose an original asynchronous Laplace transform method formalized for quasilinear problems based on the well-known Gaver-Stehfest algorithm.
Parallel experiments show the convergence of our new method, as well as several interesting properties compared with the classical algorithms.
\end{abstract}

\begin{keywords}
Laplace transform; Gaver-Stehfest algorithm; asynchronous iterations; parallel computing; quasilinear equation; option pricing 
\end{keywords}

\section{Introduction}

Laplace transforms are powerful tools employed to derive the analytical solutions of differential equations.
However, it is too difficult to find or evaluate the analytical inverse transform in closed form.
Researchers try to use numerical Laplace transform methods that convert the time-dependent equations to parameter-dependent problems, which could be parallelized easily in frequency domain.
There exist a great deal of approaches that could be used to obtain the inverse transform over the past five decades.
Davies and Martin in 1979 wrote a survey~\cite{Davies1979} investigating many promising methods, in which 14 specific algorithms were tested and compared.
Duffy~\cite{Duffy1993}  concentrated on three methods developed later, in which two methods are based on existing techniques but considerably improved by other researchers.
Cohen wrote a comprehensive review book~\cite{Cohen2007} that shows all aspects of Laplace transform inversion.
Finally, Kuhlman surveyed five different methods in~\cite{Kuhlman2013} and their implementations.

There exist many time-parallel methods, including shooting-type methods, space-time domain decomposition methods, space-time multi-grid methods, and direct methods~\cite{Gander2015}.
Obviously, Laplace transform belongs to the direct methods where~\cite{Crann1998} and~\cite{Sheen2000} are cited as pioneering works in this category.
Our paper is based on the algorithm described in~\cite{Crann1998} that is called Gaver-Stehfest algorithm, which has a good accuracy on a fairly wide range of functions as illustrated in~\cite{Davies1979}, thereafter they developed an iterative scheme in~\cite{Lai2005,Lai2010} through linearization of implied volatility, leading to a series of iterations throughout the time-stepping process.

Our purpose in this paper is to employ asynchronous iterative scheme in modeling a Laplace-type solver, which might be extremely flexible in data transmission and exploitation throughout the processing phase.
Such iterative scheme was first theoretically proposed in~\cite{Chazan1969} for relaxation algorithms, formalized further in~\cite{Miellou1975,Baudet1978,ElTarazi1982} with norm-based contraction model and in~\cite{Bertsekas1983,Bertsekas1989} with nested set model.
Another well-known time-parallel method, called Parareal, has been successfully generalized to the asynchronous iterative scheme~\cite{Magoules2018,Magoules2018d}.
In the next section, we recall and present the classical Gaver-Stehfest algorithm.
In Section~\ref{sec:il}, we introduce the iterative Laplace transform method for an option pricing model with variational volatility.
In Section~\ref{sec:async}, we propose a new Laplace method based on asynchronous iterative scheme with some remarks about the its convergence.
Finally, some numerical experiments are given in Section~\ref{sec:exp} and concluding remarks are presented in Section~\ref{sec:con}.

\section{The Gaver-Stehfest algorithm}
\label{sec:gs}

Given a second-order linear elliptic operator $\mathcal{A}$, consider the initial value problem
\begin{equation}
\label{eq:gpb}
\begin{cases}
\frac{\partial u(x,t)}{\partial t} + \mathcal{A}u(x,t) = b, & t \in [0, T],\ x\in\Omega, \\
u(x,t) = u_0(x), & t = 0,\ x \in \Omega, \\
\end{cases}
\end{equation}
where the boundary conditions are supposed defined to have a well-posed problem.
In the following we note $u=u(t)=u(\cdot,t)$ if the space variable does not hamper the formulation and computation.
To solve this equation, we employ the Laplace transform defined as
\begin{equation}
\label{eq:L}
U(z) = \int_0^\infty e^{-zt}u(t)dt.
\end{equation}
A general contour integral formula for the inverse Laplace transform is given as
\begin{equation}
\label{eq:Linv}
u(t) = \frac{1}{2\pi j}\int_\Gamma e^{zt}U(z)dz,
\end{equation}
where $\Gamma$ is the Bromwich contour that must be further determined, thus yielding a lot of numerical applications in recovering the original function.

We are interested in the Gaver-Stehfest algorithm, which aims to approximate Equation~\eqref{eq:Linv} by a sequence of functions
\begin{equation}
\label{eq:gs}
u(t) \approx \frac{\ln2}{t}\sum_{i=1}^p \omega_i U(\frac{i\ln2}{t}),
\end{equation}
where $\omega_i$ are defined as follows
\begin{equation}
\label{eq:gs-coef}
\omega_i = (-1)^{\frac{p}{2}+i}\sum_{k=\left\lfloor\frac{1+i}{2}\right\rfloor}^{\min(i, \frac{p}{2})}\frac{k^{\frac{p}{2}}(2k)!}{(\frac{p}{2}-k)!k!(k-1)!(i-k)!(2k-i)!},\quad i = 1,\dots,p,
\end{equation}
where $p$ is an even number denoting the number of processors.
This numerical solver could be deduced from Equation~\eqref{eq:Linv} by Cauchy integral formula with specific parameters.
Notice that $\omega_i$ will change sign during iterations.
Such behavior is remarkable and indeed affects the numerical performance of Gaver-Stehfest algorithm in some cases, which will be shown in the following sections, thus ignored for the moment.

\begin{figure}
\centering
\includegraphics[width=.9\linewidth]{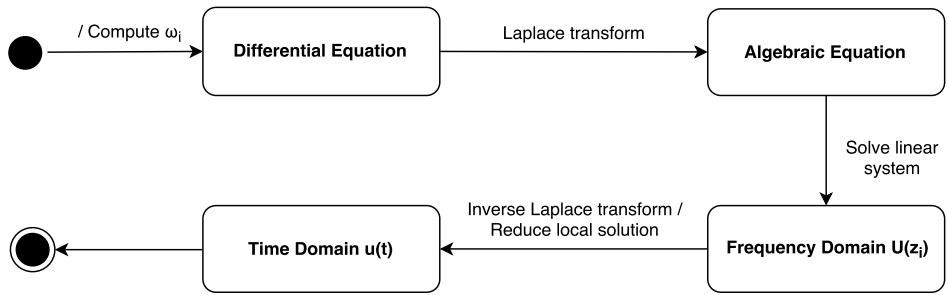}
\caption{State diagram of direct Laplace transform method.}
\label{fig:stat-lin}
\end{figure}

The implementation of direct Laplace transform method is quite simple.
Given the time-dependent problem \eqref{eq:gpb}, making Laplace transform yields
\[
z_i U(z_i) - u(0) + \mathcal{A}U(z_i) = B_i,\quad i = 1,\dots,p,
\]
where $B_i$ is the Laplace transform of $b$ depending on $z_i$, and obviously
\[
z_i = \frac{i \ln 2}{t},\quad i = 1,\dots,p.
\]
Finally, making inverse Laplace transform yields
\[
u(t) \approx \frac{\ln2}{t}\sum_{i=1}^p \omega_i U(z_i).
\]
which leads to Algorithm~\ref{alg:Llin}.
\begin{algorithm}[H]
\caption{Direct Laplace transform method for linear equation.}
\label{alg:Llin}
\begin{algorithmic}[1]
\State Compute $\omega_i$ using~\eqref{eq:gs-coef}
\State Compute $U(z_i)$ by Laplace transform
\State Reduce $\omega_i$ and $U(z_i)$ to process 1
\If{rank == 1}
\State Compute $u(t)$ using~\eqref{eq:gs}
\EndIf
\end{algorithmic}
\end{algorithm}
We notice that one should operate a reduction operation after Laplace transform, which is a commonly used term borrowed from message-passing specification, where a designated root node receives data from all nodes and executes an arithmetic or logical operation.
Here the specific operation is summation as indicated above.
Furthermore, we illustrate the corresponding state diagram in Figure~\ref{fig:stat-lin}.
Note that Laplace transform method operates in the frequency domain, whereas most of other methods are proceeding in the original space, in which time-dependent terms must be solved by appropriate temporal methods, such as Euler methods, Runge-Kutta methods, and multistep methods.

\section{Iterative Laplace transform method for quasilinear problem}
\label{sec:il}

It is not desirable to solve quasilinear problems directly and thus we deal with them in a different direction.
Throughout this paper, we take Black-Scholes equation as an example to investigate the behavior of iterative methods (see,~e.g.,~\cite{Lai2005}).

We note that option pricing is a crucial target in financial decision-making.
The breakthrough came with the appearance of the Black-Scholes equation~\cite{Black1973}, which has a huge influence to the financial market and drives an unexpected prosperity in the trading of derivatives.
This equation is often called the Black-Scholes-Merton (BSM) equation since it was further generalized by Merton in important ways~\cite{Merton1973,Merton1974}.
Consider the following European call option pricing problem
\[
\frac{\partial V(S,t)}{\partial t} + rS\frac{\partial V(S,t)}{\partial S} +
\frac{1}{2}\sigma^2S^2\frac{\partial^2 V(S,t)}{\partial S^2} = rV(S,t),
\]
where $V$ is the option price, depending on stock price $S$ and time $t$.
As before, we note $V=V(t)=V(\cdot,t)$ in the following if possible.
Volatility $\sigma$ and risk-free interest rate $r$ are the constant parameters,
with the final and boundary conditions given by
\begin{equation}
\label{eq:bs-ibv}
\begin{cases}
V(S,t) = \max(S-E, 0), & t = T,\ S \in [0, +\infty), \\
V(S,t) = 0, & t \in [0, T],\ S = 0, \\
V(S,t) \sim S - Ee^{-r(T-t)}, & t \in [0, T],\ S \rightarrow +\infty,
\end{cases}
\end{equation}
where $T$ is the maturity of the option and $E$ is the strike price.
If there exist transaction costs, then the price prediction may become much more complex.
The volatility can be treated in different ways, for instance using a modified volatility function $\tilde\sigma$.
Here we assume that the price of option is a parameter of $\tilde\sigma$, thus leading to the BSM equation with implied volatility
\begin{equation}
\label{eq:bs}
\frac{\partial V}{\partial t} + rS\frac{\partial V}{\partial S} +
\frac{1}{2}{\tilde\sigma(V)}^2S^2\frac{\partial^2 V}{\partial S^2} = rV.
\end{equation}
There are many ways to choose an appropriate $\tilde\sigma$.
Here we adopt the function described in \cite{Lai2005} as following
\begin{equation}
\label{eq:bs-sigma}
\tilde\sigma(V) = \sigma\sqrt{1 + \sin\left(\frac{\pi V}{E}\right)},
\end{equation}
where $\sigma$ denotes constant historical volatility.

We should apply Laplace transform method to Equation~\eqref{eq:bs}, but this one is too complex to be employed directly.
Moreover, from~\eqref{eq:bs-ibv} we notice that~\eqref{eq:bs} is a backward equation.
Hence, we perform the following variable transformation
\[
S = Ee^x,\quad t = T - \frac{2\tau}{\sigma^2},\quad V = Su(x, \tau),\quad \kappa = \frac{2r}{\sigma^2},
\]
substituting into Equation~\eqref{eq:bs} gives
\begin{equation}
\label{eq:bst}
\frac{\partial u}{\partial \tau} =  \frac{\tilde\sigma(u)^2}{\sigma^2}(\frac{\partial^2 u}{\partial x^2} + \frac{\partial u}{\partial x}) + \kappa\frac{\partial u}{\partial x},
\end{equation}
with the corresponding conditions
\begin{equation}
\label{eq:bst-ibv}
\begin{cases}
u(x,\tau) = \max(1-e^{-x}, 0), & \tau=0,\ x\in\mathbb{R}, \\
u(x,\tau) = 0, & \tau\in[0, \frac{T\sigma^2}{2}],\ x\rightarrow-\infty, \\
u(x,\tau) \sim 1-e^{-\kappa\tau-x}, & \tau\in[0, \frac{T\sigma^2}{2}],\ x\rightarrow+\infty.
\end{cases}
\end{equation}
Implied volatility defined in~\eqref{eq:bs-sigma} becomes
\begin{equation}
\label{eq:bst-sigma}
\tilde\sigma(u) = \sigma\sqrt{1 + \sin(\pi ue^x)}.
\end{equation}
Notice that~\eqref{eq:bst} is forward parabolic and the fractional term is flexible for both linear and quasilinear model.
If we import a constant volatility, the fractional term will vanish; if we use implied volatility like the one defined~\eqref{eq:bst-sigma}, the constant coefficient therein will also disappear.

\begin{figure}
\centering
\includegraphics[width=1.\linewidth]{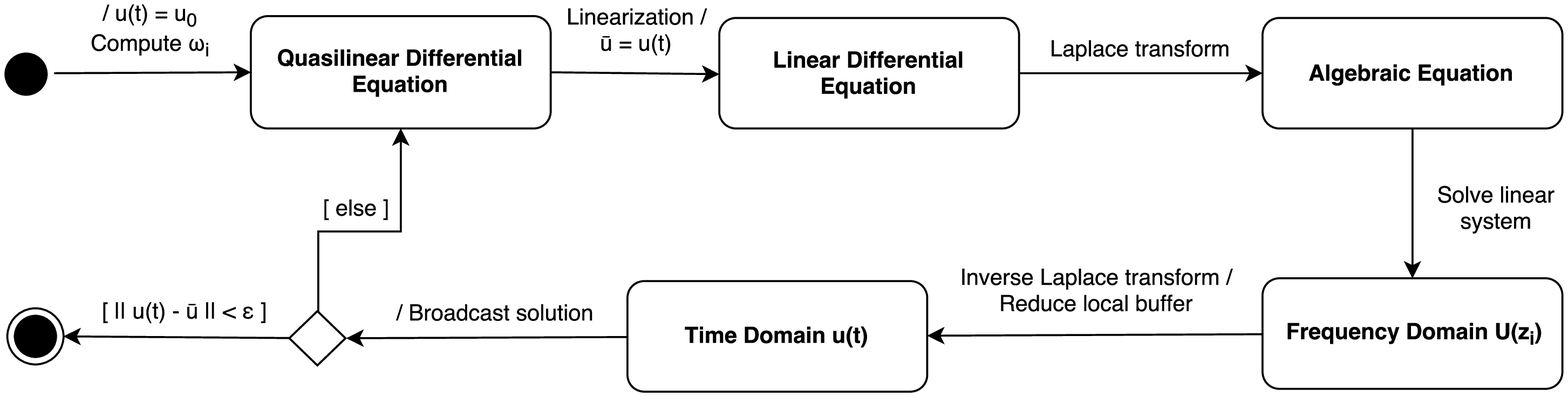}
\caption{State diagram of synchronous Laplace transform method.}
\label{fig:stat-sync}
\end{figure}

This equation, however, depends on a quasilinear coefficient that should be tackled before getting forward.
We share the idea of~\cite{Lai2005} by introducing a one-step retard into $a(u)$, which linearizes the quasilinear equation~\eqref{eq:bst} in the form
\begin{equation}
\label{eq:bst-lin}
\frac{\partial u}{\partial \tau} =  \frac{\tilde\sigma(\bar{u})^2}{\sigma^2}(\frac{\partial^2 u}{\partial x^2} + \frac{\partial u}{\partial x}) + \kappa\frac{\partial u}{\partial x}.
\end{equation}
We note that $\bar{u}$ can be seen as a ``frozen variable'' which is assigned the previous value of $u$ and thus does not depend on~$\tau$.
Now we prepare to apply the Laplace transform method to Equation~\eqref{eq:bst-lin}.
Setting
\[
a(\bar u) = \frac{\tilde\sigma(\bar{u})^2}{\sigma^2},
\]
yields
\begin{equation}
\label{eq:bst-lt}
z_i U - u(x, 0) = a(\bar{u})(\frac{\partial^2 U}{\partial x^2} + \frac{\partial U}{\partial x}) + \kappa\frac{\partial U}{\partial x}.
\end{equation}
More specifically, this yields an iterative scheme illustrated in Algorithm~\ref{alg:Lsync}.
The same algorithm has been investigated in~\cite{Lai2005} under the name of ``iterative coefficient--inverse Laplace transform''.
\begin{algorithm}[H]
\caption{Synchronous Laplace transform method for quasilinear equation.}
\label{alg:Lsync}
\begin{algorithmic}[1]
\State $u(t) = u_0$
\State Compute $\omega_i$ using~\eqref{eq:gs-coef}
\Repeat
\State $\bar{u} = u(t)$
\State Compute $U(z_i)$ by Laplace transform
\State Reduce $\omega_i$ and $U(z_i)$ to process 1
\If{rank == 1}
\State Compute $u(t)$ using~\eqref{eq:gs}
\EndIf
\State Broadcast $u(t)$ to other processes
\Until{$\norm{u(t) - \bar{u}} \simeq 0$}
\end{algorithmic}
\end{algorithm}
Since all processors need the values of $u(t)$ at the beginning of each iteration, we should broadcast such buffer to all neighbors.
Note that ``broadcast'' is a collective communication operation which involves a specific root node communicating its data to all other nodes.
We observe that $\omega_i$ is computed outside because it depends only on the rank $i$ and the number of processors $p$.
Similarly, the state diagram is shown in Figure~\ref{fig:stat-sync}.

We mention here that there are large numbers of models dealing with the transaction cost, such as those in~\cite{Boyle1992} and~\cite{Barles1998}, some of which are nonlinear differential equations that have different forms and properties.
There exist meanwhile many other types of equations related to the option pricing problem, such as the martingale pricing theory~\cite{Harrison1979}, the binomial options pricing model~\cite{Cox1979}, and the stochastic volatility option models~\cite{Hull1987,Duan1995}.
They are completely different from the Black-Scholes model and thus should be addressed separately.
In general, one might employ Algorithm~\ref{alg:Lsync} to other types of equations generalized from Equation~\eqref{eq:gpb}, such as
\[
\frac{\partial u}{\partial t} + a(u)\mathcal{A}u = b.
\]
Substituting the frozen variable and applying Laplace transform yields
\[
z_i U - u(0) + a(\bar{u})\mathcal{A}U = B_i,
\]
where linear operator $\mathcal{A}$ can be solved by any appropriate discretization methods.

\section{Asynchronous Laplace transform method}
\label{sec:async}

\subsection{Formalization}

Asynchronous iterative method releases the restriction of strict data dependency whereby iterations are executed by several processors in arbitrary order without any synchronization during computation, first formalized in \cite{Chazan1969} for linear systems.
Obviously, it must be subject to some constraints that each processor still needs to continuously obtain the latest information.
Since there is no synchronization imposed to such scheme, asynchronous iterative algorithm may exhibit incredible flexibility and efficiency even with the expansion of computational nodes, which contributes to overcome the fault tolerance and the load balancing problems.

\begin{figure}
\centering
\includegraphics[width=.7\linewidth]{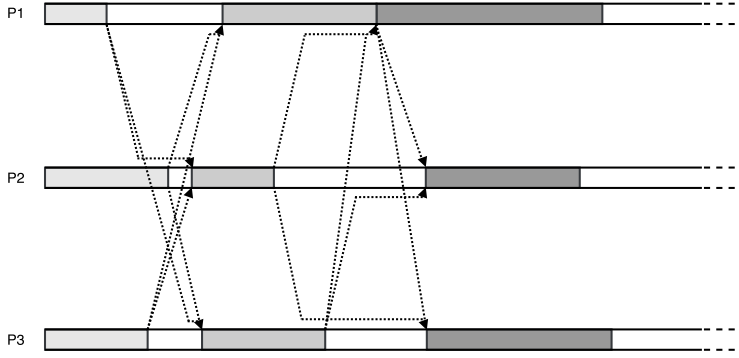}
\caption{Synchronous iterations with asynchronous communications.}
\label{fig:siac}
\end{figure}

Consider a vector space $E$ with
\[
E = E_1 \times\dots\times E_p.
\]
Let
\[
f:E\rightarrow E,\quad f_i:E\rightarrow E_i,\quad i\in\{1,\dots,p\}.
\]
Practically, $p$ denotes the number of processors.
Hence
\[
f(x) = \left[f_1(x_1,\dots,x_p)\ \dots\ f_p(x_1,\dots,x_p)\right]^\intercal,\quad x = \left[x_1\ \dots\ x_p\right]^\intercal,\quad x\in E,
\]
leading to the classical parallel iterative scheme
\begin{equation}
\label{eq:sync}
x_i^{k+1} = f_i\left(x_1^k,\dots,x_p^k\right),\quad i\in\{1,\dots,p\},
\end{equation}
which is presented in Figure~\ref{fig:siac}.
In order to compute $(k+1)$th iteration's $x$ in processor $i$, one needs to collect all data of iteration $k$ from other processors, which imposes a synchronization point on the computational framework at the end of each iteration.

\begin{figure}
\centering
\includegraphics[width=.7\linewidth]{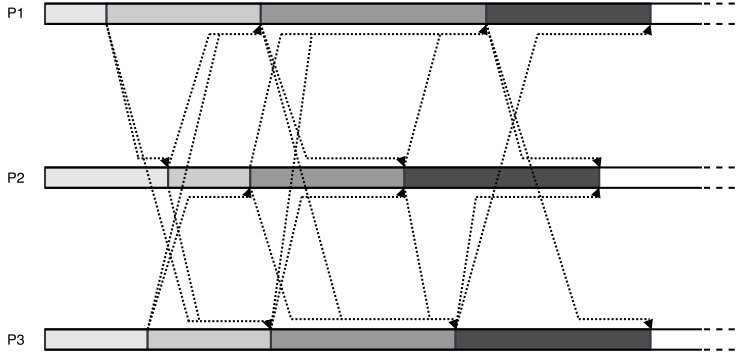}
\caption{Asynchronous iterations with asynchronous communications.}
\label{fig:aiac}
\end{figure}

Now we define the integer subsets $\left\{P^{(k)}\right\}_{k\in\mathbb{N}}$ such that
\[
P^{(k)} \subset \{1,\dots,p\},\quad P^{(k)}\ne\emptyset,\quad \forall k\in\mathbb{N},
\]
and let $\rho_{i,j}(k)$ be nonnegative integer-valued functions satisfying
\begin{equation}
\label{eq:conr}
\rho_{i,j}(k) \le k,\quad k\in\mathbb{N},
\end{equation}
with $i,j\in\{1,\dots,p\}$.
We introduce here an iterative scheme that generates the following sequence
\begin{equation}
\label{eq:async}
x_i^{k+1} =
\begin{cases}
f_i\left(x_1^{\rho_{i,1}(k)},\dots,x_p^{\rho_{i,p}(k)}\right), & i \in P^{(k)}, \\
x_i^k, & i \notin P^{(k)},
\end{cases}
\end{equation}
which is actually the mathematical model of asynchronous iterative scheme and illustrated in Figure~\ref{fig:aiac}.
Generally, for the sake of convergence analysis, Assumption~\ref{ass:async:conr} and Assumption~\ref{ass:async:conp} are armed therewith, which ensure that processors read eventually the latest information for each element from essential neighbors and no processor stops updating during iterations.
\begin{assumption}
\label{ass:async:conr}
$\forall i,j\in\{1,\dots,p\},\quad \underset{k \to +\infty}{\lim}\rho_{i,j}(k) = +\infty$.
\end{assumption}
\begin{assumption}
\label{ass:async:conp}
$\forall i\in\{1,\dots,p\},\quad \text{card}\left\{k\in\mathbb{N}\mid i\in P^{(k)}\right\} = +\infty$.
\end{assumption}

\begin{figure}
\centering
\includegraphics[width=1.\linewidth]{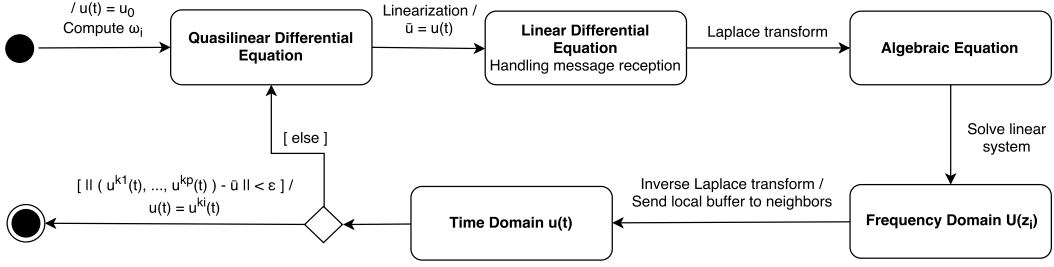}
\caption{State diagram of asynchronous Laplace transform method.}
\label{fig:stat-async}
\end{figure}

Now we apply asynchronous iterative scheme to the iterative Laplace transform method depicted in Algorithm~\ref{alg:Lsync}.
Since newer value $u(t)$ depends on the previous value $\bar{u}$, thereafter we use $u^{k+1}$ and $u^k$ to denote these two values during iterations.
let
\begin{equation}
\label{eq:opL}
\mathcal{L}_i\left(u^k\right) = U(z_i),\quad i\in\{1,\dots,p\},\quad k\in\mathbb{N},
\end{equation}
and for a fixed time span $t$, we define
\begin{equation}
\label{eq:opG}
u_i^{k+1} = \mathcal{G}_i(U) = \frac{\ln2}{t}\omega_i U(z_i),\quad i\in\{1,\dots,p\},\quad k\in\mathbb{N}.
\end{equation}
Notice that
\[
u^{k+1} = \frac{\ln2}{t}\sum_{i=1}^p \omega_i U(z_i) = \sum_{i=1}^p u_i^{k+1},\quad k\in\mathbb{N},
\]
which leads to a handy operator
\begin{equation}
\label{eq:opS}
u^k = \mathcal{S}\left(u_1^k,\dots,u_p^k\right) = \sum_{i=1}^p u_i^k,\quad k\in\mathbb{N}.
\end{equation}
Combining~\eqref{eq:opL}, \eqref{eq:opG}, and~\eqref{eq:opS} gives
\[
u_i^{k+1} = \left(\mathcal{L}_i \circ \mathcal{G}_i \circ \mathcal{S}\right)\left(u_1^k,\dots,u_p^k\right),\quad i\in\{1,\dots,p\},\quad k\in\mathbb{N}.
\]
Setting
\begin{equation}\label{eq:fi}
f_i = \mathcal{L}_i \circ \mathcal{G}_i \circ \mathcal{S},\quad i\in\{1,\dots,p\},
\end{equation}
yields
\[
u_i^{k+1} = f_i\left(u_1^k,\dots,u_p^k\right),\quad i\in\{1,\dots,p\},\quad k\in\mathbb{N},
\]
which follows exactly the classical parallel iterative scheme \eqref{eq:sync}.
Hence, iterative Laplace method can be generalized naturally to the asynchronous iterative scheme subject to \eqref{eq:conr} and assumed to satisfy Assumption~\ref{ass:async:conr} and Assumption~\ref{ass:async:conp}
\begin{equation}
\label{eq:Lasync}
u_i^{k+1} =
\begin{cases}
f_i\left(u_1^{\rho_{i,1}(k)},\dots,u_p^{\rho_{i,p}(k)}\right), & i \in P^{(k)}, \\
u_i^k, & i \notin P^{(k)},
\end{cases}
\end{equation}
then leading to Algorithm~\ref{alg:Lasync}.
\begin{algorithm}[H]
\caption{Asynchronous Laplace transform method for quasilinear equation.}
\label{alg:Lasync}
\begin{algorithmic}[1]
\State $u^0(t) = u_0$
\State Compute $\omega_i$ using~\eqref{eq:gs-coef}
\State $k_i = 0$
\Repeat
\State $k_i = k_i + 1$
\State $\bar{u}_i = u^{k_i-1}(t)$
\ForAll{ $j\in\{1,\dots,i-1,i+1\dots,p\}$ }
\State Request receiving $U^{\rho_{i,j}(k_j)}(z_i)$ from process $j$
\EndFor
\State Compute $U^{k_i}(z_i)$ by Laplace transform
\ForAll{ $j\in\{1,\dots,i-1,i+1\dots,p\}$ }
\State Request sending $U^{k_i}(z_i)$ to process $j$
\EndFor
\State Compute $u^{k_i}(t)$ using~\eqref{eq:gs}
\Until{$\norm{(u^{k_1}(t),\dots,u^{k_p}(t)) - (\bar{u}_1,\dots,\bar{u}_p)} \simeq 0$}
\State $u(t) = u^{k_i}(t)$
\end{algorithmic}
\end{algorithm}
The most remarkable notation therein is $k_i$ that exhibits the chaotic behavior of asynchronous iterative scheme, where intrinsically retard term $\rho_{i,j}$ and execution set $P^{(k)}$ play important roles during iterations.
Meanwhile, another hinge consists in the sending and receiving operations without synchronization point, which yields the asynchronous performance, as opposed to the aforementioned blocking reduction and broadcast operations.
Note that one may also employ nonblocking reduction and broadcast operations to implement Algorithm~\ref{alg:Lasync}.
The state diagram is depicted in Figure~\ref{fig:stat-async}.

\subsection{Remarks on convergence}

To the best of our knowledge, unfortunately, there is no available theory that can be used to prove the convergence of asynchronous Laplace algorithm.
According to the norm-based convergence theory (see~\cite{ElTarazi1982}, which is based on the results in~\cite{Miellou1975,Baudet1978}), one should prove a relationship in the form
\[
\norm{f(x)-f(y)}_\infty^w \le \alpha\norm{x-y}_\infty^w,\quad \forall x,y\in E,
\]
where $\alpha\in(0,1)$ and $\norm{.}_\infty^w$ denotes the weighted maximum norm with $w>0$, in order to establish the convergence result.
This imposes a contraction condition to the operator $f$. 
We recall that in our case there exists a summation operator $\mathcal{S}$, as seen in~\eqref{eq:fi}, and thus $f$ is not an assembly but a summation of subvectors.
It is clear that the summation operator could not lead to desired partial order under contraction.
A more general but similar idea proposed in~\cite{Bertsekas1983} (see also~\cite{Bertsekas1989}) is based on the nested set theory, in which a technique known as ``box condition'' allows to obtain convergence result without using any norm.
This can be written as
\[
E^k = E_1^k \times \cdots \times E_p^k,\quad k\in\mathbb{N},
\]
where $E_i^k \subset E_i$ for all $i$.
This condition involves a superposition of infinite subsets and thus suffers from the same problem as the previous approach.
In addition, two-stage and nonstationary models asynchronous models were discussed in~\cite{Frommer1994,Frommer2000} and some papers made use of the partial ordering such as~\cite{ElBaz1996b,Miellou1998}.
These models have gained much popularity in some problems but do not correspond to our case.

The expansion of summation operator is indeed compensated by the frequency decomposition, which may contribute to the convergence.
In what follows we will not pursue the convergence issue further; instead, we will focus on the numerical behavior of our method.

\section{Numerical experiments}
\label{sec:exp}

\subsection{Environmental setting}

We have conducted several experiments for the direct Laplace transform method, using the Gaver-Stehfest algorithm as illustrated in Algorithm~\ref{alg:Llin}, as well as the iterative variants based on Algorithms~\ref{alg:Lsync} and~\ref{alg:Lasync}.
All tests are executed using~\eqref{eq:bst-lt} with $\tilde{\sigma}(\bar{u})=\sigma$ in the linear case, whereas \eqref{eq:bst-sigma} is adopted in the quasilinear case.

All tests are launched on an SGI ICE X cluster connected with InfiniBand, involving MPI library to run parallel applications, which is supported by SGI-MPT 2.14.
Mathematical operations and linear systems solvers are implemented by the Alinea programming library~\cite{Magoules2015} and iterative algorithms are programmed by JACK~\cite{Magoules2017b,Magoules2018b} for both synchronous and asynchronous variants.

\subsection{Direct scheme for linear equation}

\begin{figure}[!t]
\centering
\begin{subfigure}{.50\textwidth}
  \centering
  \includegraphics[width=1.\linewidth]{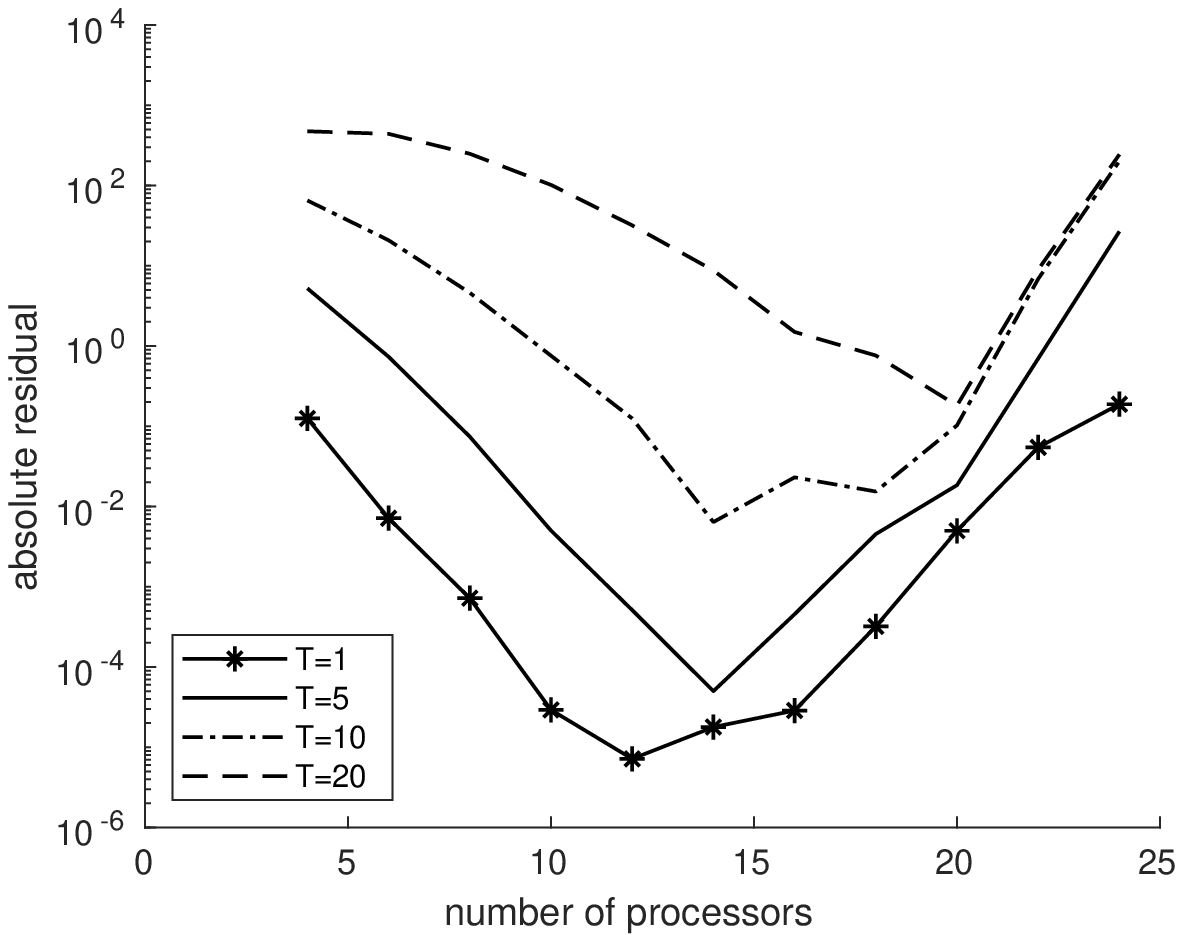}
\end{subfigure}\begin{subfigure}{.50\textwidth}
  \centering
  \includegraphics[width=1.\linewidth]{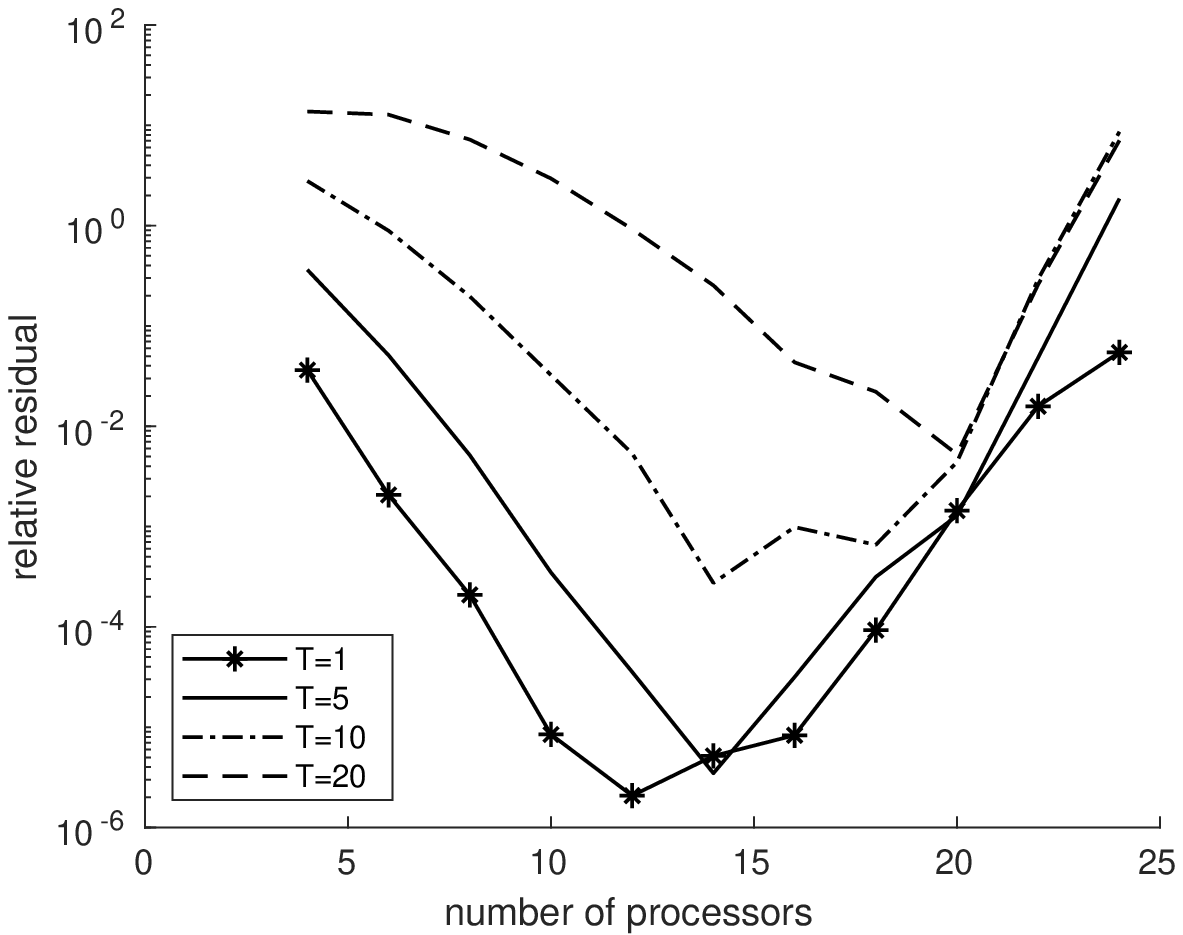}
\end{subfigure}
\caption{Direct Laplace transform method for linear equation: absolute error (left), relative error (right).}\label{fig:Llin}
\end{figure}

According to \cite{Stehfest1970}, number of processors $p$ must be even.
We first illustrate the numerical results of Algorithm~\ref{alg:Llin} applied to linear equation, shown in Figure~\ref{fig:Llin}.
Here we run programs with different numbers of processes and observe that error behaves like a convex function for each specific time span, where $T$ denoting the maturity of the specific option contract.
We choose $10^{-3}$ as convergence threshold, which is adequate and conspicuous for the financial data.
Table~\ref{tab:Llin-1} gives the convergence interval of direct method.
\begin{table*}[t]
\caption{Convergence zone of direct Laplace transform method with large maturities.\label{tab:Llin-1}}
\centering
\begin{tabular}{lcc}
\toprule
$T$ & convergence interval of $p$ & remarks \\
\midrule
1 & 8, 10, 12, 14, 16, 18 & $p<8$: inaccurate; $p>18$: inaccurate \\
5 & 10, 12, 14, 16, 18 & $p<10$: inaccurate; $p>18$: inaccurate \\
10 & 14, 16, 18 & $p<14$: inaccurate; $p>18$: inaccurate \\
20 & $\emptyset$ & inaccurate, oscillating \\
\bottomrule
\end{tabular}
\end{table*}
Here, convergence interval and convergence zone are defined as the set of $p$ that leads to convergent result.
Notice that the larger we assign for time span, the narrower convergence interval we get.
When $T=20$, there is no convergence interval observed throughout experiments.
In this case, we should compute results step-by-step, as described in~\cite{Lai2005}.
Here ``oscillating'' indicates that we could not observe a contracting behavior when $p$ increases monotonously.
On the contrary, results fall into an inconsistent solution range and oscillate divergently around initial point.
We mention here that large $T$, denoting the maturity with the unit of year, is not a normal test case in option pricing problem, which is shown only for the experimental purpose.
In the following tests, we will use primarily small $T$ to illustrate the real behaviors.

Davies and Martin~\cite{Davies1979} mentioned that the Laplace transform method based on the Gaver-Stehfest algorithm gives good accuracy on a fairly wide range of functions.
Figure~\ref{fig:Llin} and Table~\ref{tab:Llin-1} show that a well-posed option pricing problem leads to good accuracy with a fairly wide range of convergence interval (e.g., $T=1$ lead to a convergence interval $8\le p\le18$).
Theoretically, the more processors we use, the more precision we obtain through parallel processing~\cite{Kuznetsov2013}.
In practice, however, we must consider the arithmetical precision of the test machine, namely, we decrease the truncation error thanks to the large number of processors and meanwhile increase the rounding error due to digit width limit.
Unfortunately, the side effect is dramatically large due to the rapid growth of $\omega_i$ with sign alternating.
Thus we could not exploit the power of large scale parallel computing, which acts as an intrinsic defect of the Gaver-Stehfest algorithm.

\subsection{Iterative scheme for quasilinear equation}

Now we focus on the numerical behavior of iterative Laplace transform methods illustrated in Algorithms~\ref{alg:Lsync} and~\ref{alg:Lasync}.
We continue to use the convergence threshold $10^{-3}$ as the termination condition, which may lead to an infinite loop in our program.
Experimental results are shown in Tables~\ref{tab:Lsync} and~\ref{tab:Lasync}.
\begin{table*}[t]
\caption{Convergence zone of synchronous Laplace transform method.\label{tab:Lsync}}
\centering
\begin{tabular}{lcc}
\toprule
$T$ & convergence interval of $p$ & remarks \\
\midrule
0.01 & 6, 8, 10, 12 & $p<6$: inaccurate; $p>12$: divergent \\
0.1 & 4, 6, 8, 10, 12 & $p<4$: inaccurate; $p>12$: divergent \\
1 & 4, 6, 8, 10, 12 & $p<4$: inaccurate; $p>12$: divergent \\
\bottomrule
\end{tabular}
\end{table*}
\begin{table*}[t]
\caption{Convergence zone of asynchronous Laplace transform method.\label{tab:Lasync}}
\centering
\begin{tabular}{lcc}
\toprule
$T$ & convergence interval of $p$ & remarks \\
\midrule
0.01 & 6 & $p<6$: inaccurate; $p>6$: divergent \\
0.1 & 6 & $p<6$: inaccurate; $p>6$: divergent \\
1 & 4, 6 & $p<4$: inaccurate; $p>6$: divergent \\
\bottomrule
\end{tabular}
\end{table*}
We notice that the convergence interval of the asynchronous Laplace transform method is much narrower than the synchronous method.
This is a reasonable compensation because asynchronous methods require less waiting time in communication but more computational iterations than classical parallel scheme, which amplify the unstable issue of $\omega_i$ with rapid growth and sign alternating.

On the other hand, for a large $p$, both synchronous and asynchronous algorithms become divergent.
This means that we could not obtain a result underneath the given threshold since the uncertainty propagates throughout iterations.
Hence, such behavior is caused by an intrinsic issue of the Gaver-Stehfest algorithm, which is the same as that in the direct method.

We illustrate in Table~\ref{tab:Llin-2} the results of direct Laplace method with the same interval of $T$ as in Tables~\ref{tab:Lsync} and~\ref{tab:Lasync}.
\begin{table*}[t]
\caption{Convergence zone of direct Laplace transform method with small maturities.\label{tab:Llin-2}}
\centering
\begin{tabular}{lcc}
\toprule
$T$ & convergence interval of $p$ & remarks \\
\midrule
0.01 & 4, 6, 8, 10, 12, 14, 16, 18, 20, 22 & $p<4$: inaccurate; $p>22$: inaccurate \\
0.1 & 6, 8, 10, 12, 14, 16, 18, 20 & $p<6$: inaccurate; $p>20$: inaccurate \\
1 & 6, 8, 10, 12, 14, 16, 18 & $p<6$: inaccurate; $p>18$: inaccurate \\
\bottomrule
\end{tabular}
\end{table*}
One should keep in mind that the iterative results are obtained from the BSM equation with implied volatility, whereas the direct method, as seen in Algorithm~\ref{alg:Lsync}, is applied to the linear equation where $\tilde\sigma=\sigma$ is a constant.
Thus, comparing the former with the latter could not lead to an obvious conclusion for the performance of our method.
We show these results only for clarity and completeness.

\begin{figure}[!t]
\centering
\begin{subfigure}{.48\textwidth}
  \centering
  \includegraphics[width=1.\linewidth]{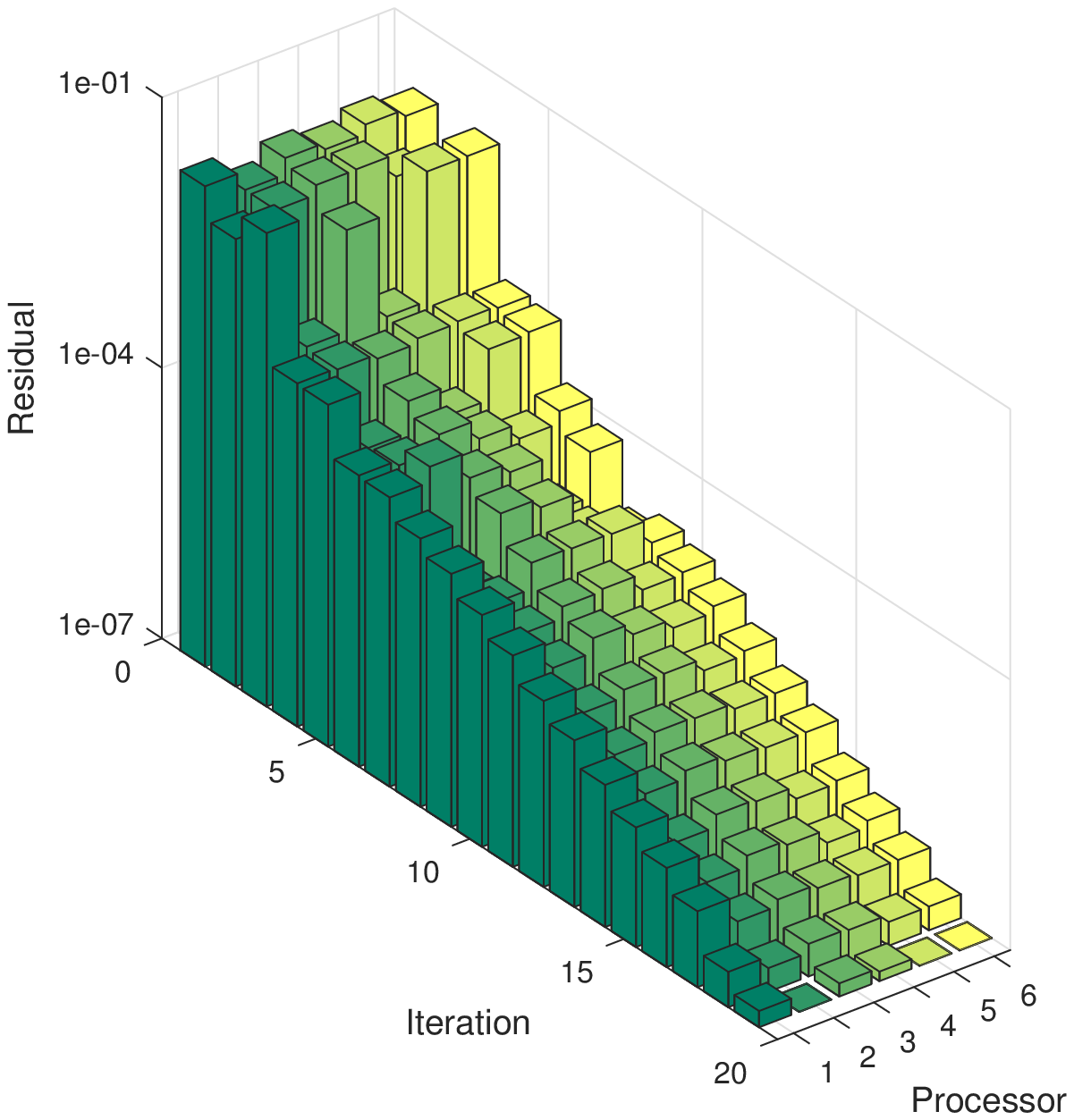}
\end{subfigure}\begin{subfigure}{.52\textwidth}
  \centering
  \includegraphics[width=1.\linewidth]{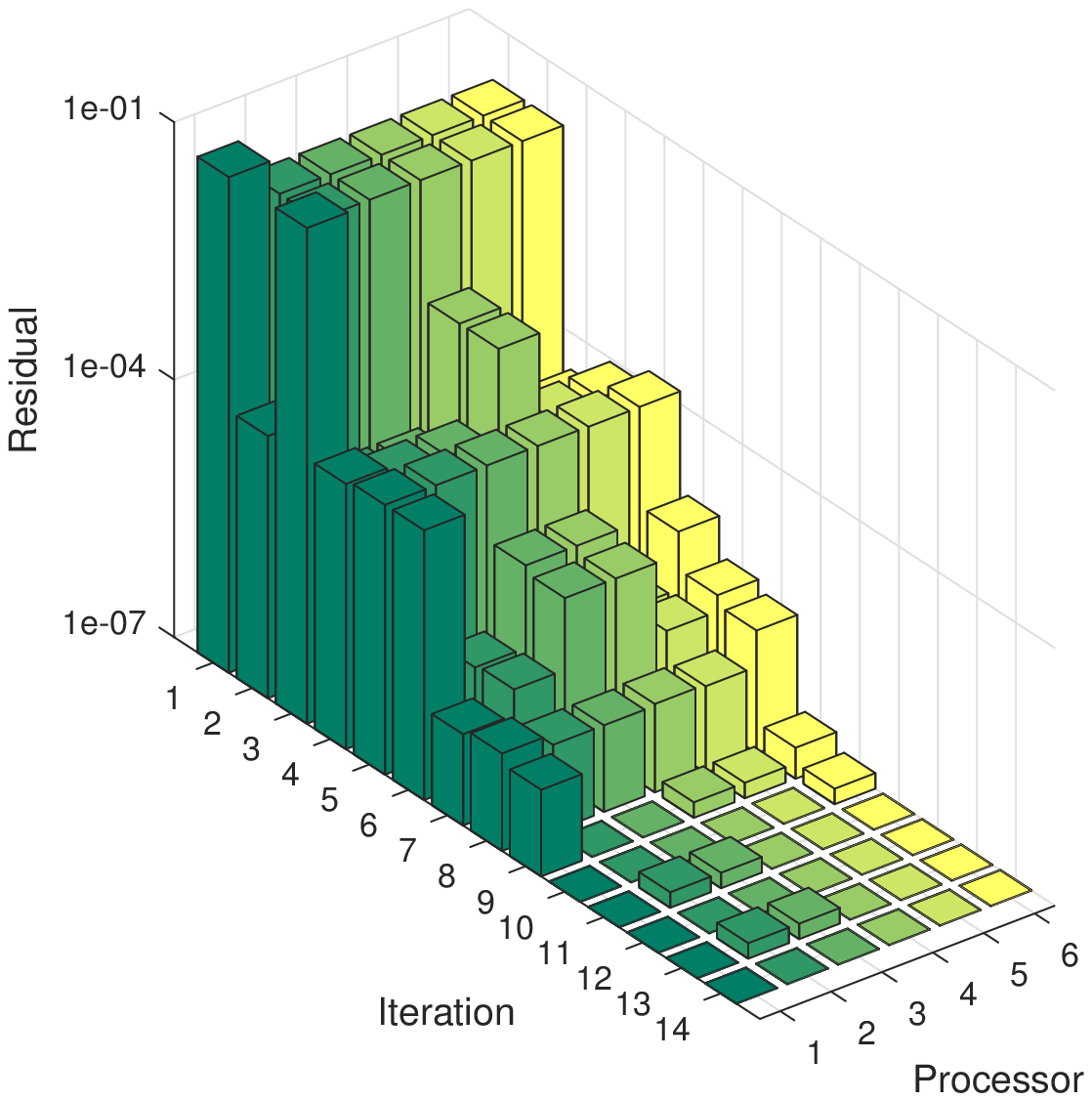}
\end{subfigure}
\caption{An example of successive asynchronous Laplace iteration with $p=6$, $\Delta T=0.1$, $n=10$: first time interval (left), last time interval (right).}\label{fig:Lasync}
\end{figure}

Figure~\ref{fig:Lasync} illustrates another experiment that divides temporal interval $T$ into several smaller parts, as described in~\cite{Lai2005}.
As mentioned before, we address primarily the real situation in option pricing problem where $T$ is small.
For a small $T$ we need no more subdivision in the time domain.
Thus, we conduct this test only for evaluating the performance of the algorithm illustrated in~\cite{Lai2005}, where we execute asynchronous Laplace transform method step-by-step along time domain.
Similar to the classical temporal integration schemes like Euler methods, we perform computations in each step using always the latest data, but employ a rather wide subinterval to fit the characteristic of the Laplace transform method.
The number of processors, the step size, and the number of steps are chosen as $p=6$, $\Delta T=0.1$, $n=10$, respectively.
We can see that the execution process in the first time interval is highly structured, except for some unordered data due to the asynchronous scheme.
Nonetheless, the last time interval gives different behavior that seems somewhat affected by the unfinished chores due to the asynchronous iterative scheme.
This produces larger retards and converges more quickly when receiving the latest data in some steps.

Finally, we test the accuracy of the asynchronous Laplace method in comparison with the synchronous version.
We fix the volatility $\sigma=0.3$, risk-free rate $r=0.05$, maturity $T=1$, and number of processor $p=6$.
Recall that
\[
\tilde\sigma(V) = \sigma\sqrt{1 + \sin\left(\frac{\pi V}{E}\right)}.
\]
The constant volatility and the strike price in the right hand side are fixed, whereas the volatility $\tilde{\sigma}$ changes with $V$.
Then we change the stoke price $S$ and strike price $E$ to compare the two methods.
Results are shown in Table~\ref{tab:price}.
\begin{table*}[t]
\caption{Synchronous and asynchronous Laplace results with synchronous option prices $V_\text{sync}$, asynchronous option prices $V_\text{async}$, absolute errors $\varepsilon_\text{abs}$, and relative errors $\varepsilon_\text{rel}$, given $\sigma=0.3$, $r=0.05$, $T=1$, and $p=6$.\label{tab:price}}
\centering
\begin{tabular}{llcccc}
\toprule
$S$ & $E$ & $V_\text{sync}$ & $V_\text{async}$ & $\varepsilon_\text{abs}$ & $\varepsilon_\text{rel}$ \\
\midrule
$60$ & $50$ & $32.178769$ & $32.178726$ & $4.3*10^{-5}$ & $1.34 * 10^{-6}$ \\
$100$ & $50$ & $66.517320$ & $66.517268$ & $5.2*10^{-5}$ & $7.82 * 10^{-7}$ \\
$20$ & $30$ & $7.797859$ & $7.797825$ & $3.4*10^{-5}$ & $4.36 * 10^{-6}$ \\
$20$ & $50$ & $5.380797$ & $5.380790$ & $7.0*10^{-6}$ & $1.30 * 10^{-6}$ \\
\bottomrule
\end{tabular}
\end{table*}
Here, $V_\text{sync}$ and $V_\text{async}$ are the option prices obtained from synchronous and asynchronous iterations, respectively.
$\varepsilon_\text{abs}$ and $\varepsilon_\text{rel}$ act as absolute and relative errors such that
\[
\varepsilon_\text{abs} = \abs{V_\text{sync}-V_\text{async}},\quad \varepsilon_\text{rel} = \frac{\varepsilon_\text{abs}}{V_\text{sync}}.
\]
We notice that the asynchronous option prices are close enough to the synchronous cases.
This observation achieves our expectations, as we speculated that the chaotic iterative process does not affect dramatically the final result.
Since the synchronous Laplace transform method has proved to be accurate in predicting the option price values using the BSM equation with implied volatility, we can draw a conclusion that our method yields accurate results as well.

\subsection{Further discussion}

We have seen that Laplace transform methods are highly influenced by the number of digits of machine precision that are prone to round-off error propagation, which leads to an unstable behavior.
Notice that~\eqref{eq:gs-coef} increases quickly with commutative plus-minus sign.
The accuracy mounts in the beginning as the number of processes $p$ increases and then decreases rapidly.

One could avoid such effect for Algorithm~\ref{alg:Llin} by simply adopting the recommended precision $\mu_{GS} = 1.1 * p$ in a multi-precision computational environment as discussed in~\cite{Abate2004}.
Meanwhile, the fixed Talbot method~\cite{Talbot1979} requires the precision $\mu_{FT} = p$, which performs a less restrictive behavior and seems a better alternative to the Gaver-Stehfest algorithm according to the experiments therein.
Further investigation gives a unified framework for various types of inverse Laplace transform algorithms~\cite{Abate2006}.
Recall that the contour integral is
\[
u(t) = \frac{1}{2\pi j}\int_\Gamma e^{zt}U(z)dz.
\]
Let $y = zt$, we have
\[
u(t) = \frac{1}{2\pi jt}\int_\Gamma e^{y}U(\frac{y}{t})dy.
\]
Then taking
\[
e^y \approx \sum_{i=0}^n \frac{\beta_i}{\alpha_i - y},
\]
yields
\[
u(t) \approx \frac{1}{t}\sum_{i=0}^n \frac{1}{2\pi j}\int_\Gamma \frac{\beta_i}{\alpha_i - y}U(\frac{y}{t})dy.
\]
Finally, using Cauchy integral formula, this becomes
\begin{equation}
\label{eq:Luni}
u(t) \approx \frac{1}{t}\sum_{i=0}^n \beta_i U(\frac{\alpha_i}{t}).
\end{equation}
Choosing
\[
p = n,\quad \alpha_i = \ln 2,\quad \beta_i = \omega_i\ln 2,
\]
where $p$ is an even integer, and notice that $\beta_0 = 0$, we get again the Gaver-Stehfest formula \eqref{eq:gs}.
We could further rewrite~\eqref{eq:Luni} in the form
\[
u_i^{k+1} = \frac{1}{t}\beta_i U(\frac{\alpha_i}{t}),\quad i\in\{1,\dots,n+1\},\quad k\in\mathbb{N},
\]
where
\[
U(\frac{\alpha_i}{t}) = \mathcal{L}_i\left(u^k\right),\quad i\in\{1,\dots,n+1\},\quad k\in\mathbb{N}.
\]
Combining~\eqref{eq:opL} and~\eqref{eq:opS}, we could obtain the asynchronous scheme~\eqref{eq:Lasync} for the unified Laplace transform model, whereby the methods discussed in~\cite{Abate2006}, such as Fourier series methods and Talbot method, might also be executed in asynchronous mode, although no rigorous proofs are available.
The contour integral approach with quadrature is another choice and has proved to be very efficient for parabolic problems with time-independent coefficients~\cite{Sheen2000}.
One may expect to combine these strategies to tackle the quasilinear equations in a more effective way.

These issues are beyond the scope of this paper.
We leave this work, as well as a rigorous proof of the asynchronous convergence, as a direction of future investigation.

\section{Concluding remarks}
\label{sec:con}

In this paper, we investigated the Laplace transform and proposed a new method based on the asynchronous iterative scheme.
We formalized the mathematical model and gave some remarks on its convergence.
Numerical experiments showed that asynchronous Laplace transform method converges chaotically to the correct solution sets.
The intrinsic numerical approximation problem affects the accuracy of Gaver-Stehfest algorithm and led to a narrow convergence interval.
There exists much remains to be done to prove the convergence of this method, which might require a complete extension of the existing asynchronous theory.
Further investigation could focus on the asynchronous formalization for other efficient inversion formulas and nonlinear solvers.

\bibliography{ref}
\bibliographystyle{abbrv}

\end{document}